# Phonon-pump electronic-probe study of methylammonium lead iodide reveals electronically decoupled organic and inorganic sublattices

Peijun Guo[1], Arun Mannodi-Kanakkithodi[1], Jue Gong[2], Yi Xia[1], Constantinos C. Stoumpos[3], Duyen H. Cao[4], Benjamin T. Diroll[1], John B. Ketterson[5], Gary P. Wiederrecht[1], Tao Xu[2], Maria K. Y. Chan[1], Mercouri G. Kanatzidis[3], Richard D. Schaller[1,3,*]

[1]Center for Nanoscale Materials, Argonne National Laboratory, 9700 South Cass Avenue, Lemont, IL 60439, United States

[2]Department of Chemistry and Biochemistry, Northern Illinois University, 1425 W. Lincoln Hwy., DeKalb, IL 60115, United States

[3]Department of Chemistry, Northwestern University, 2145 Sheridan Road, Evanston, IL 60208, United States

[4]Materials Science Division, Argonne National Laboratory, 9700 South Cass Avenue, Lemont, IL 60439, United States

[5]Department of Physics and Astronomy, Northwestern University, 2145 Sheridan Road, Evanston, IL 60208, United States

**Abstract**

**Organic-inorganic hybrid perovskites such as methylammonium lead iodide ($CH_3NH_3PbI_3$) are game-changing semiconductors for solar cells[1] and light-emitting devices[2] owing to their exceptionally long carrier lifetime and diffusion length[3]. Determining whether the large dipole moment of the organic cation and dynamic disorder benefit the optoelectronic properties of $CH_3NH_3PbI_3$ has been an outstanding challenge[4, 5]. Herein, via transient absorption measurements employing an infrared pump pulse tuned to a methylammonium vibration, we observe slow, nanosecond-long thermal dissipation from the selectively excited organic mode to the entire lattice. Resulting transient electronic signatures, during the period of thermal-nonequilibrium when the induced thermal motions are mostly concentrated on the organic sublattice, reveal that induced motions of the organic cations do not alter absorption or photoluminescence response of $CH_3NH_3PbI_3$, beyond thermal effects. Our results suggest that the attractive optoelectronic properties of $CH_3NH_3PbI_3$ mainly derive from the inorganic lead-halide framework[6-8].**

Hybrid perovskites, including the prototypical $CH_3NH_3PbI_3$, present challenges in understanding the interplay between structural and optoelectronic properties. This primarily stems from significant dynamic disorder of the organic cations ($CH_3NH_3^+$) with picosecond, observed relaxation times[4, 9]. Due to crucial importance for enhancing device efficiency[10] and enabling new functionalities[11, 12], the mutual interactions between the fluctuating dipolar organic cations and charge carriers that reside on the inorganic Pb-I framework have been extensively investigated. Formation of (anti)ferroelectric domains arising from polar organic cations was proposed to enhance carrier lifetimes[13], but spectroscopic measurement did not support such a picture[14]. Dynamic bulk Rashba effect induced by the $CH_3NH_3^+$ cations was also suggested[15, 16], yet recent theoretical and experimental results show that such an effect is negligible[17]. In fact, $HC(NH_2)_2PbI_3$-based devices exhibit conversion efficiencies comparable to the $CH_3NH_3PbI_3$ analog[18], albeit the dipole moment of $HC(NH_2)_2^+$ is much smaller than that of $CH_3NH_3^+$ (ref[13]). Although comparative studies between hybrid and all-inorganic perovskites have been attempted so as to evaluate the role of polar organic cations on material performance[19-21], the properties of metal-halide perovskites are known to depend on synthetic conditions and material history[22, 23]. Experimental approaches permitting direct investigation of interactions between the organic and inorganic sublattices in hybrid systems are thereby important.

Here, we employ infrared-pump electronic-probe (IPEP) spectroscopy, to excite strongly-absorbing vibrational modes of the organic sublattice and examine the corresponding optical response near the bandgap of $CH_3NH_3PbI_3$. Fig. 1a maps the temperature dependent static absorption of a $CH_3NH_3PbI_3$ thin film around its bandgap. The tetragonal-to-orthorhombic phase transition near 145 K is denoted by an abrupt change of bandgap, and for each individual phase the bandgap increases with temperature. The near-bandgap, distinct absorption peak in the orthorhombic phase arises due to a comparatively stronger excitonic character. Fig. 1b presents temperature-dependent static absorption arising from N-H asymmetric stretching modes (abbreviated here as N-H-*as* modes) of the $CH_3NH_3^+$ cation, which exhibit the strongest absorption cross-section among various vibrational modes of the molecular cation[24]. A dramatic increase of the N-H-*as* absorption accompanying the phase transition is due to the formation of hydrogen bonding that suppresses rotational motions of $CH_3NH_3^+$ in the orthorhombic phase[25]. The pronounced mid-infrared absorption, in conjunction with the near-bandgap absorption feature, facilitate infrared-pump electronic-probe experiments on the orthorhombic phase of $CH_3NH_3PbI_3$.

In transient absorption measurements, we selectively pumped the N-H-*as* modes (centered at 3120 cm$^{-1}$; Fig. S1) and probed the change of optical density ($\Delta OD$) near the bandgap. The $\Delta OD$ spectral map acquired on $CH_3NH_3PbI_3$ at 10 K is displayed in Fig. 1c. We observe a derivative-like transient spectral response, where magnitude following excitation starts from zero, grows with pump-probe time delay, and saturates on the nanosecond timescale. Positive $\Delta OD$ signal on the blue side of the static absorption peak and negative signal to the red conveys a spectral blueshift, consistent with transient lattice heating (Fig. 1a). Note that large bandgap shifts arising from the optical Stark effect were observed at zero delay time, when the pump and probe pulses temporally overlapped[26] (Fig. S2). An off-resonance pump (centered at 2200 cm$^{-1}$) results in negligible transient response (Fig. S3), confirming that the response shown in Fig. 1c is due to vibrational excitation of the organic sublattice. The nanosecond timescale relates slow down-conversion of thermal energy from the high-energy N-H-*as* modes to the low-lying, cold phonon bath, with the latter encompassing all $CH_3NH_3^+$ modes except the N-H-*as* modes, and all the modes of the inorganic sublattice.

Probing the tetragonal crystal phase at 205 K distills a comparably faster thermal equilibration with a timescale of hundreds of picoseconds (Fig. S4). Measurement near the phase transition (~145 K) reveals that the two coexisting phases respond independently to the pump (Fig. 1d), as no evidence of transient phase change is observed, which would otherwise lead to the transfer of spectral weight between the two phases. The $\Delta OD$ amplitude near the bandgap obtained for the tetragonal phase is, as expected, weaker than that for the orthorhombic counterpart, owing to a weaker vibrational absorption (Fig. 1b). The nearly identical timescales of $\Delta OD$ kinetics for the co-existing phases at 140 K (Fig. S4d) indicate that the rotations of $CH_3NH_3^+$, which are only present in the tetragonal phase, do not apparently influence phonon-phonon coupling strength between the two sublattices. Temperature-dependent kinetics (Fig. 2a) present slower thermal equilibration at lower temperatures. Furthermore, the $\Delta OD$ amplitude increases with decreasing temperature (Fig. 2b), which results from the heat capacity that diminishes, and the absorption cross-section of N-H-*as* modes that grows, with decreasing temperature. Fluence-dependent measurements (Fig. 2c) show that the $\Delta OD$ amplitude scales nearly linearly with pump fluence, and the thermal equilibration time is not sensitive to pump fluence over the explored range, consistent with earlier work based on transient mid-infrared spectroscopy using electronic excitation and an infrared probe[24].

Having established the transient response with IPEP, we then turn to investigating whether perturbatively induced motions of the organic sublattice influence the near-bandgap optical properties of $CH_3NH_3PbI_3$. As illustrated in Fig. 3a, following the selective excitation of N-H-*as* modes, phonon-phonon scattering leads to the sequential population of various lower-energy $CH_3NH_3^+$ vibrational modes, and ultimately the population of inorganic phonon modes. Note that thermal equilibration within the inorganic sublattice itself requires only 10~20 ps, as reported elsewhere[6]. Therefore, the nanosecond timescale is mainly attributed to (1) slow intramolecular vibrational relaxation (IVR) within the $CH_3NH_3^+$ cations[27], supported by the relatively small phonon scattering phase space located in the 200~3000 cm$^{-1}$ range (Fig. 3a); (2) slow energy transfer from $CH_3NH_3^+$ to the inorganic framework[24]. To gain more insights into the vibrational evolution, we performed ab-initio molecular dynamics simulations on structures with selective excitation of the N-H-*as* modes based on the phonon calculations (see Supplementary Note 1), from which we extracted atomic trajectories up to ~130 ps. As shown in Fig. 3b, selective excitation of N-H-*as* modes results in the population of stretching modes as represented by the variation of bond lengths of N-H, C-H and C-N. The C-N bond-length variation decays more slowly in comparison to the C-H and N-H bonds, as the latter two are associated with vibrational modes of highest frequencies and depopulate most rapidly via IVR. Note that librational motions, captured by the orientational variation of the C-N bond (Fig. 3c), can also be induced. The simulated timescale is furthermore not strongly dependent on the total imparted energy in the simulations, which is consistent with the experimental results (Fig. 2c).

The change in amplitude of transient optical signal can act as a proxy for how much thermal energy has been transferred to the inorganic framework (that in turn determines the bandgap). The transient response at long delay time (>1000 ps), when the sample has reached thermal equilibrium (Fig. 2a), results from the change of bandgap with temperature. Transient spectra plotted for representative delay times (Fig. 3d) during the thermal relaxation closely overlap with a simple scaling of amplitude. Such spectral similarity suggests that no net increase or decrease of optical absorption coefficient, other than that caused by lattice heating, occurs upon this perturbation when $CH_3NH_3PbI_3$ is at *thermal non-equilibrium* (i.e., 0~1000 ps). Because optical absorption can be viewed as an inverse process of bimolecular recombination[28], Fig. 3d implicitly suggests that pump-induced $CH_3NH_3^+$ motions do not impact bimolecular recombination in $CH_3NH_3PbI_3$.

IPEP measurements on $CH_3NH_3PbBr_3$ reveal similar transient optical response (Fig. S5) compared to $CH_3NH_3PbI_3$ due to their structural similarity. However, in contrast to $CH_3NH_3PbI_3$ and $CH_3NH_3PbBr_3$ which are three-dimensional perovskites, impact of artificially induced organic cation motion is observed for the two-dimensional perovskite, $(CH_3(CH_2)_3NH_3)_2PbI_4$. For this compound[29], the $CH_3(CH_2)_3NH_3^+$ cations situated between the perovskite layers strongly impact the electronic properties through both quantum and dielectric confinement effects. Following selective vibrational excitation near 3100 cm$^{-1}$ (Fig. S6), a much faster transient response, relative to the 3D counterparts, was observed for $(CH_3(CH_2)_3NH_3)_2PbI_4$ near the exciton energy (Fig. S5), owing to IVR processes of the $CH_3(CH_2)_3NH_3^+$ cations that in turn alters the strength of quantum confinement of the electron-hole pairs (i.e., excitons) residing on the Pb-I framework.

We further explored if the infrared-pump-induced $CH_3NH_3^+$ motions can directly impact photoluminescence (PL) of $CH_3NH_3PbI_3$, using infrared-pump PL-probe measurements based on the schematic shown in Fig. S7. Here, the pump pulse was followed by a time delayed (denoted as $\Delta t$), low-intensity, 400-nm PL-probe pulse that generates electron-hole pairs (positive $\Delta t$ is defined if the infrared pump hits the sample first). The emitted photons were either detected by a CCD camera, or spectrally and temporally resolved by a streak camera. Fig. 4a presents a representative, spectrally-resolved PL map measured at 50 K without using the infrared pump. Fig. 4b shows the spectrally-integrated PL decay kinetics measured under various negative values of $\Delta t$, with the infrared pump turned on. The dip in the PL decay trace, which exhibits an instrument-response-time-limited timescale, is also observed by using an off-resonant mid-infrared pump (Fig. S8), and therefore attributed to electric-field driven dissociation of electron-hole pairs, or the optical Stark effect. We then focused on measurements under positive $\Delta t$ values. Fig. 4c presents PL decays measured using different positive $\Delta t$ values that traverse the duration of sublattice thermal-equilibration. We find that optically pumping the organic vibrational modes of $CH_3NH_3PbI_3$ does not alter the carrier recombination rate.

While streak camera measurements only detect PL over specific time windows, time-integrated PL measurements with higher signal-to-noise ratio can reveal whether the infrared-pump-induced $CH_3NH_3^+$ motions can enhance (by suppressing defect scattering of charge carriers) or suppress (by slowing down the recombination rate of carriers so that more trapping events can take place) the brightness of $CH_3NH_3PbI_3$. Static, temperature dependent PL intensity (Fig. 4d) suggests that a substantial portion of photo-generated carriers are still lost through non-radiative decay pathways at 50 K. The time-integrated PL spectra measured at 10 K and 50 K are shown in Fig. 4e; note that the dramatic change in PL linewidth upon temperature reduction has been observed previously[30]. For both temperatures, the infrared pump ($\Delta t$=2000 ps) yields a blueshift of the PL spectrum, stemming from a bandgap increase due to lattice heating. We ran measurements at these two temperatures because the fast PL decay component due to bimolecular recombination[31] is commensurate with the thermal equilibration time, and hence the impact of non-equilibrium $CH_3NH_3^+$ motions, if there are any, can be more pronounced. Based on the PL spectral shift at 50 K (in conjunction with data in Fig. 1a and heat capacity[32]), we estimate that 0.06 photons are absorbed per orthorhombic unit cell by $CH_3NH_3^+$, with a photon energy of 386 meV (corresponding to 3120 cm$^{-1}$). We denote the change of PL counts at each wavelength as $\Delta N = N_{on} - N_{off}$, where $N_{on}$ ($N_{off}$) designates PL counts obtained with (without) the infrared pump. The spectra of $\Delta N/I_{tot}$ measured at 10 K and 50 K using variable positive $\Delta t$ values are presented in Fig. 4f and 4g, respectively, where $I_{tot}$ is the spectrally-integrated total PL counts measured without the infrared pump. At 10 K, infrared pumping yields a PL blueshift, the magnitude of which increases with $\Delta t$. This arises since with a larger $\Delta t$, more photons were emitted subsequent to more substantial sample heating, which produces a stronger bandgap increase. A similar PL blueshift is observed also at 50 K, although an additional increase of PL intensity, implied by an asymmetric line-shape, is observed especially for small $\Delta t$ values. To explore the origin of such PL enhancement, we ran control experiments with an off-resonance infrared pump centered at 2200 cm$^{-1}$. As shown in Fig. 4h, although the PL blueshift is no longer present due to the lack of lattice heating, a PL intensity enhancement is still recovered. The observed PL enhancement retained using both pumping wavelengths is likely attributable to polaron formation arising from below-bandgap excitation of various phonon modes of the inorganic sublattice (evident from Fig. S9)[8], that may contribute to protection of charge carriers from defect scattering. This effect is not clearly observable at 10 K since the $\Delta N/I_{tot}$ intensity arising from bandgap blueshift is an order of magnitude larger than in the 50-K measurement. Importantly, comparison of Fig. 4g and 4h reveals that the induced $CH_3NH_3^+$ motions alone leads to negligible variation of the integrated PL intensity.

Our transient absorption and PL measurements both suggest that infrared-pump induced $CH_3NH_3^+$ motions do not lead to noticeable changes of the near-bandgap absorption and PL properties, beyond effects due to lattice heating. Such an observation implies that photo-generated carriers residing on the inorganic sublattice are essentially decoupled from the dynamically disordered organic cations. This conclusion is consistent with the long carrier lifetimes (exceeding 10 µs) of high-quality, fully-inorganic perovskite $CsPbI_3$, and remarkable solar cell efficiencies derived from such[33, 34]. The novel spectroscopic technique reported here, complementary to fully electronic pump-probe measurements[35, 36], can be generalized for studying organic semiconductors, organic-inorganic interfaces, and other organic-inorganic hybrid materials such as superatomic solids[37] and two-dimensional hybrid perovskites[29].

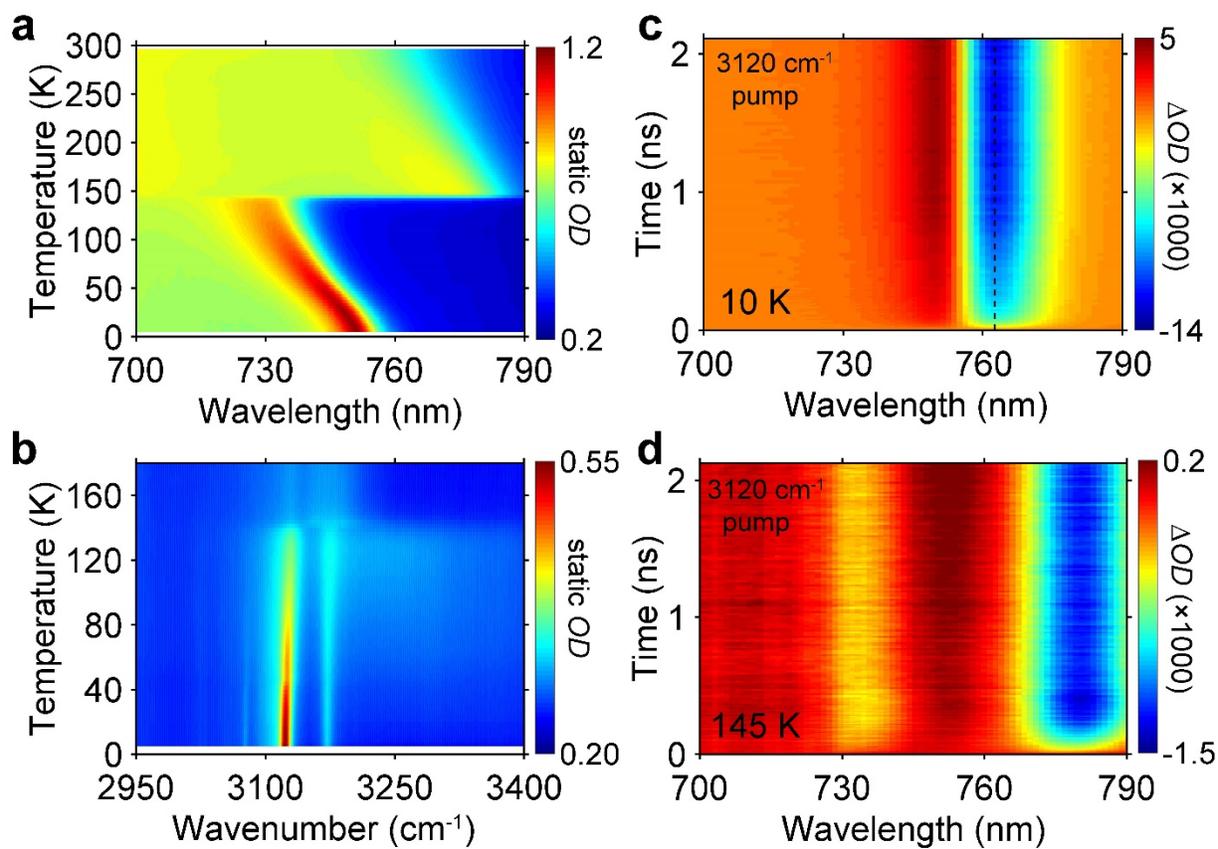

**Figure 1. Static and IPEP measurements of $CH_3NH_3PbI_3$. a** Static, temperature-dependent absorption around the bandgap. **b** Static, temperature-dependent absorption around the N-H-*as* vibrational modes. **c** $\Delta OD$ spectral map measured with on-resonance infrared pump (centered at 3120 cm$^{-1}$) at 10 K. The dashed line shows the wavelength at which the kinetics are extracted (Fig. 2a). **d** $\Delta OD$ spectral map measured at 145 K with on-resonance infrared pump. Fluence in **c** and **d** was 0.45 mJ·cm$^{-2}$.

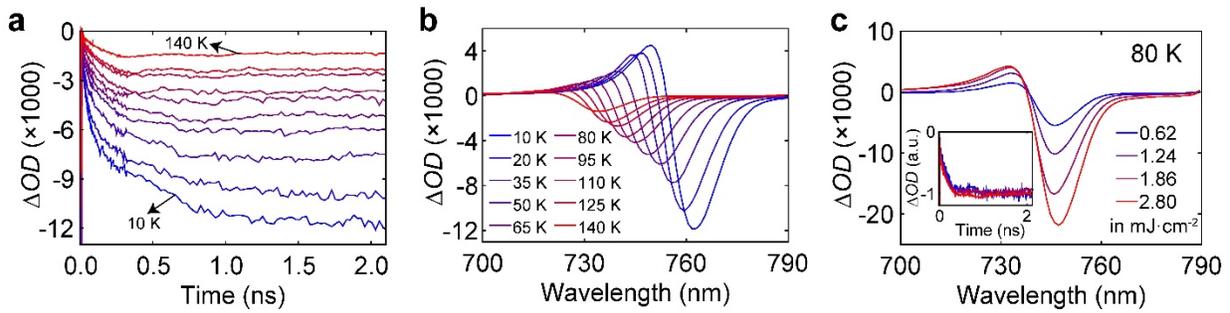

**Figure 2. Temperature and fluence dependent transient optical response. a** Transient kinetics at the wavelength indicated by the dashed line in Fig. 1c, measured at different temperatures. **b** Transient spectra at 2-ns delay time measured at different temperatures. Legend in **b** applies to **a** as well. Pump fluence used in **a** and **b** was 0.45 mJ·cm$^{-2}$. **c** Fluence-dependent transient optical spectra at 2-ns delay time; inset: kinetics for different fluences appear nearly unchanged.

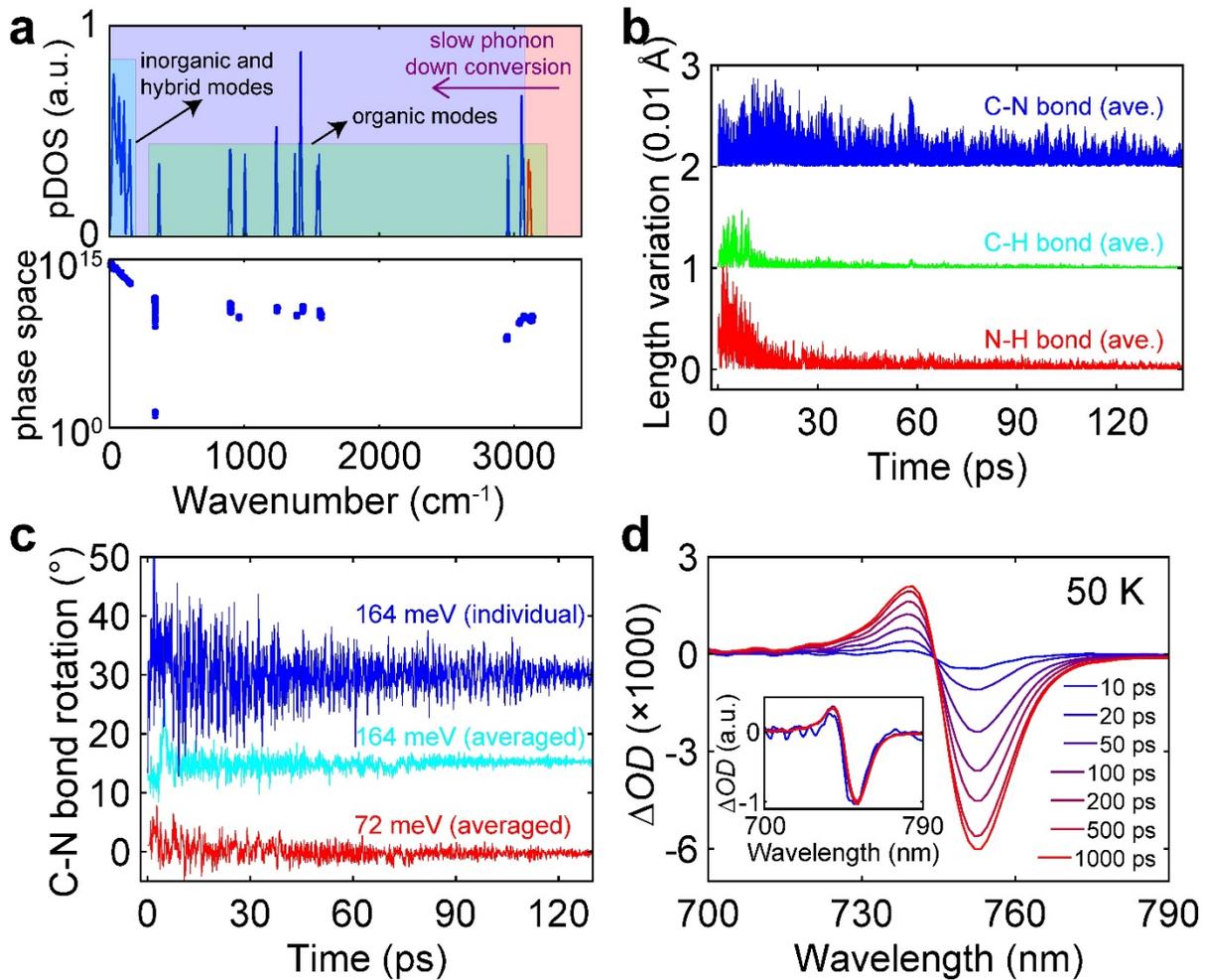

**Figure 3. First-principles calculation of phonon dynamics for orthorhombic CH₃NH₃PbI₃. a** Calculated zero-Kelvin phonon density of states (pDOS) and phonon scattering phase space. **b** Absolute variation in the N-H, C-H and C-N bond lengths from AIMD simulations. Data are averaged over the four $CH_3NH_3^+$ cations in the orthorhombic unit cell. Initial imparted energy is 164 meV. **c** Variation in the C-N bond angle for different amounts of energies imparted into the unit cell. Blue curve is for a representative C-N bond; cyan and red curves are data averaged over all the four $CH_3NH_3^+$ cations in the orthorhombic unit cell. Data in **b** and **c** are offset for clarity. **d** Transient spectra acquired at several different delay times; inset shows the overlapped transient spectra in arbitrary units by simple scaling. Fluence used was 0.45 mJ·cm⁻².

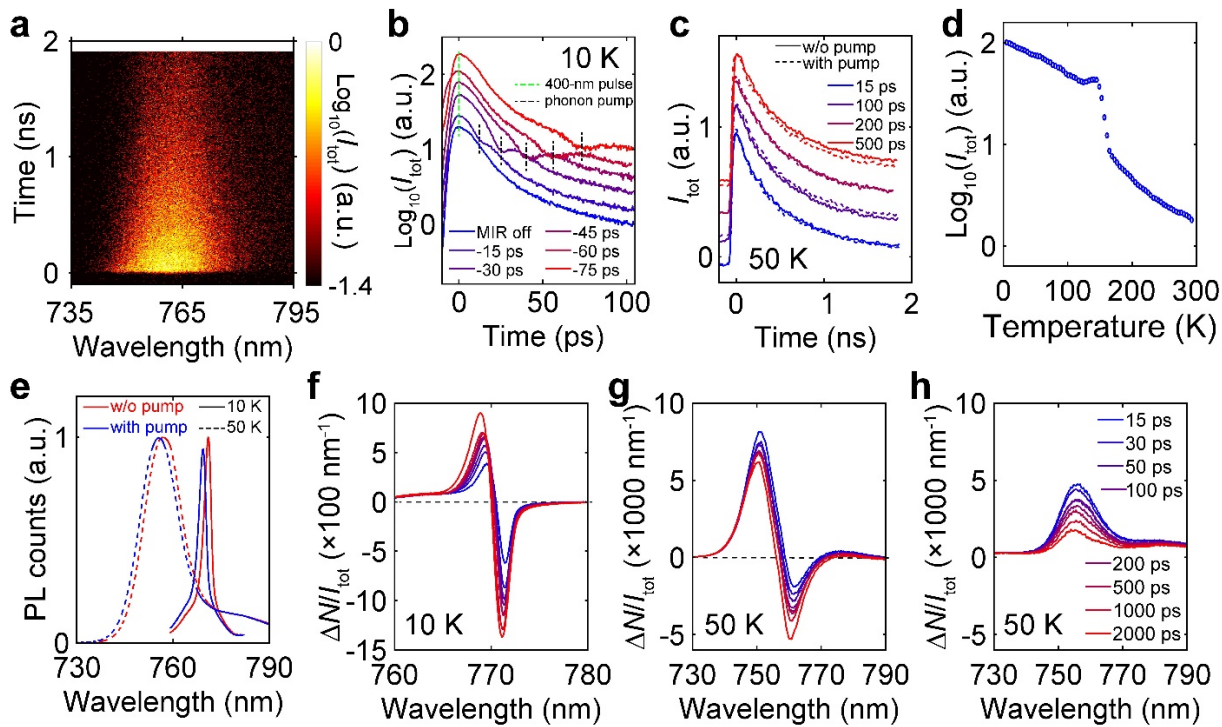

**Figure 4. Infrared-pump PL-probe measurements of CH$_3$NH$_3$PbI$_3$. a** Spectrally and temporally resolved PL counts measured with a streak camera; infrared pump was turned off. **b** PL intensity decay kinetics acquired under various negative $\Delta t$ values. **c** Comparison of PL intensity decay kinetics acquired under various positive $\Delta t$ values, with (dashed lines) and without (solid lines) using the infrared pump. **d** Static, temperature-dependent, integrated PL intensity of the measured CH$_3$NH$_3$PbI$_3$ film. **e** Time-integrated PL spectra acquired at 10 K and at 50 K, with $\Delta t$=2000 ps. Relative change of the time-integrated PL spectra with various positive delay times measured at 10 K in **f**, and measured at 50 K in **g** and **h**. On-resonance infrared pump was used in **e**, **f**, **g**; off-resonance (2200 cm$^{-1}$) infrared pump was used in **h**. Fluence of the infrared pump was fixed at 3.8 mJ·cm$^{-2}$ in all the measurements. Legend in **h** also applies to **f** and **g**.

**Methods**

*Chemicals and CH$_3$NH$_3$I synthesis.* Methylamine solution (CH$_3$NH$_2$, 40 wt. % in H$_2$O), N-methyl-2-pyrrolidinone (NMP) and γ-Butyrolactone (GBL, ≥99%) were purchased from Aldrich. Hydriodic acid (HI, 57 wt. % in H$_2$O) and lead(II) iodide (PbI$_2$, 99.9985% metals basis) were purchased from Alfa Aesar. Ethyl ether (anhydrous) and acetone were obtained from Fisher Chemical. All chemicals were used as received without further purification. Equimolar of HI was dropwise added to CH$_3$NH$_2$ with stirring in a 50-mL round bottom flask immersed in ice bath, followed by rotary evaporation at 60 ˚C to dry off the solvent. Next, white solid was obtained and washed with excessive ethyl ether on filter paper, as accompanied with vacuum filtration. Washed CH$_3$NH$_3$I powder was then dried in a vacuum oven at 60 ˚C overnight to yield final product.

*CH$_3$NH$_3$PbI$_3$ thin film fabrication.* The preparation of CH$_3$NH$_3$PbI$_3$ thin film follows method reported previously with slight modification[38]. 1.2:1 molar ratio of CH$_3$NH$_3$I/PbI$_2$ was dissolved in 1:1 volume ratio of NMP/GBL to make a 40 wt. % precursor solution. Then, precursor solution was spun-coated on an acetone-cleaned sapphire substrate with a speed of 3000 rpm for 30 s. Next, as-formed wet film was quickly transferred into an ethyl ether bath (50 mL) for 90 s. Finally, developed film was dried in air, covered by a petri dish, and followed by annealing at 150 ˚C for 15 mins on a hot plate in a humidity-controlled environment.

*Static optical characterization*

Static infrared absorbance spectra were captured by FTIR (Thermo Nicolet 6700). Static visible absorbance spectra were obtained with a customized setup. In all the optical measurements (both static and transient), the samples were mounted in a 4-K closed-cycle cold-finger cryostat under with a base pressure below 1×10$^{-7}$ Torr.

*Transient absorption measurement*

Transient absorption measurements were performed using a 35-fs amplified titanium:sapphire laser operating at 800 nm at a repetition rate of 2 kHz. Broadband visible probe pulses were generated by focusing a portion of the amplifier output into an Al$_2$O$_3$ window. Infrared pump pulses were generated by difference frequency mixing of signal and idler beams using an optical parametric amplifier and were reduced in repetition rate to 1 kHz. The probe pulses were mechanically time delayed using a translation stage and retroreflector. Full transient spectral maps are shown in Fig. S10 and S11.

*Infrared-pump PL-probe measurement*

Generation of the infrared pump pulse was identical to that employed in the transient absorption measurements. 400-nm PL excitation pulses were produced by frequency-doubling of the 800-nm amplifier output with a BBO crystal and underwent similar beam path as the visible probe pulses in the transient absorption measurements. Both the infrared pump pulses and the 400-nm PL excitation pulses were maintained at 2 kHz. Temporally and spectrally resolved PL data were collected with a streak camera. Time-integrated PL data were captured with a CCD camera. The spot size of the 400-nm PL excitation pulses was adjusted to be smaller than that of the infrared pump pulses. The fluence of 400-nm PL excitation pulse was kept below 1 µJ·cm$^{-2}$.

## Additional Information

Supplementary information is available in the online version of the paper. Correspondence and requests for materials should be addressed to R.D.S. (schaller@anl.gov, schaller@northwestern.edu).## Author contributions

P.G. conceived the infrared-pump electronic-probe measurements under the supervision of R.D.S.. J.G., C.C.S., D.H.C. and B.T.D. fabricated the samples under the supervision of T.X., M.K.G. and R.D.S.. A.M.-K. and Y.X. performed first-principles calculations under the supervision of M.K.Y.C.. G.P.W. and J.B.K. contributed to the experiments and analysis. P.G. wrote the manuscript with input from all authors. R.D.S. supervised the project.## Notes

The authors declare no competing financial interests.

## Acknowledgements

The work was performed at the Center for Nanoscale Materials, a U.S. Department of Energy Office of Science User Facility, and supported by the U.S. Department of Energy, Office of Science, under Contract No. DE-AC02-06CH11357. This material is based upon work supported by Laboratory Directed Research and Development (LDRD) funding from Argonne National13

Laboratory, provided by the Director, Office of Science, of the U.S. Department of Energy under contract DE-AC02-06CH11357. Work at Northwestern University was supported by grant SC0012541 from the US Department of Energy, Office of Science (sample preparation). T.X. acknowledges the financial support from National Science Foundation (DMR-1806152). We thank Dr. A. B. F. Martinson for comments on the paper.